\documentclass{PoS}
\usepackage{bm}
\usepackage{pifont}
\usepackage{tikz}
\usetikzlibrary{shapes,shadows,arrows}
\usepackage[utf8]{inputenc}

\title{A QUDA-branch to compute disconnected diagrams in GPUs}

\ShortTitle{A QUDA-branch to compute disconnected diagrams in GPUs}

\author{Constantia  Alexandrou\\%\thanks{speaker}\\
Department of Physics, University of Cyprus, P.O. Box 20537, 1678 Nicosia, Cyprus, and\\  
Computation-based Science and Technology Research Center, Cyprus Institute, 20 Kavafi Str., Nicosia 2121, Cyprus \\  
E-mail: \email{alexand@ucy.ac.cy}}
\author{Kyriacos Hadjiyiannakou\\
Department of Physics, University of Cyprus, P.O. Box 20537, 1678 Nicosia, Cyprus\\  
E-mail: \email{hadjigiannakou.kyriacos@ucy.ac.cy}}
\author{Giannis Koutsou\\
Computation-based Science and Technology Research Center, Cyprus Institute, 20 Kavafi Str., Nicosia 2121, Cyprus\\
E-mail: \email{g.koutsou@cyi.ac.cy}}
\author{Alexei Strelchenko\\
Scientific Computing Division, Fermilab, Batavia, IL 60510-5011, USA\\
E-mail: \email{astrel@fnal.gov}}
\author{\speaker{Alejandro Vaquero Avilés-Casco} \\
Computation-based Science and Technology Research Center, Cyprus Institute, 20 Kavafi Str., Nicosia 2121, Cyprus\\
E-mail: \email{a.vaquero@cyi.ac.cy}}

\abstract{Although QUDA allows for an efficient computation of many QCD quantities, it is
surprinsingly lacking tools to evaluate disconnected diagrams, for which GPUs are specially
well suited. We aim to fill this gap by creating our own branch of QUDA, which includes new
kernels and functions required to calculate fermion loops using several methods and fermionic
regularizations.}

\FullConference{31st International Symposium on Lattice Field Theory - LATTICE 2013\\
		July 29 - August 3, 2013\\
		Mainz, Germany}

\begin{document}

\section{Introduction}

Graphics Processing Units (GPUs) are swiftly changing the panorama of lattice QCD, due to
the unprecedented increase of computer power provided by the graphic cards. Nonetheless, lack
of efficient libraries for specific task usually prevents scientist from benefiting from the
raw processing power the GPUs offer. A great effort has been done in this direction with the
nVidia-based QUDA library \cite{Clark:2009wm,Babich:2011np}, featuring many tools to generate
and analyze configurations.

Although the QUDA library is fairly complete at the present moment, it is still under development,
and new capabilities are added every day. One of the capabilities we were missing was specific
code for calculating disconnected diagrams; the present work is devoted to fill this empty space.

\section{Integrating the different methods in QUDA}
In order to calculate disconnected diagrams, we mainly need to solve the equation

\begin{equation}
M\left|s_r\right\rangle = \left|\eta_r\right\rangle,
\end{equation}
but QUDA already has a large set of inverters. The previous step of generating the random
source is done at this moment in CPUs, using external libraries for high quality random number
generation (although inclusion in GPUs is planned in the near future). In this section we will
mostly deal with the specific variation reduction techniques, adapted for GPU computation.

\subsection{Implementation of the Truncated Solver Method}

\begin{figure}[h!]
\begin{center}
$\begin{array}{cc}
\includegraphics[width=0.42\textwidth,angle=0]{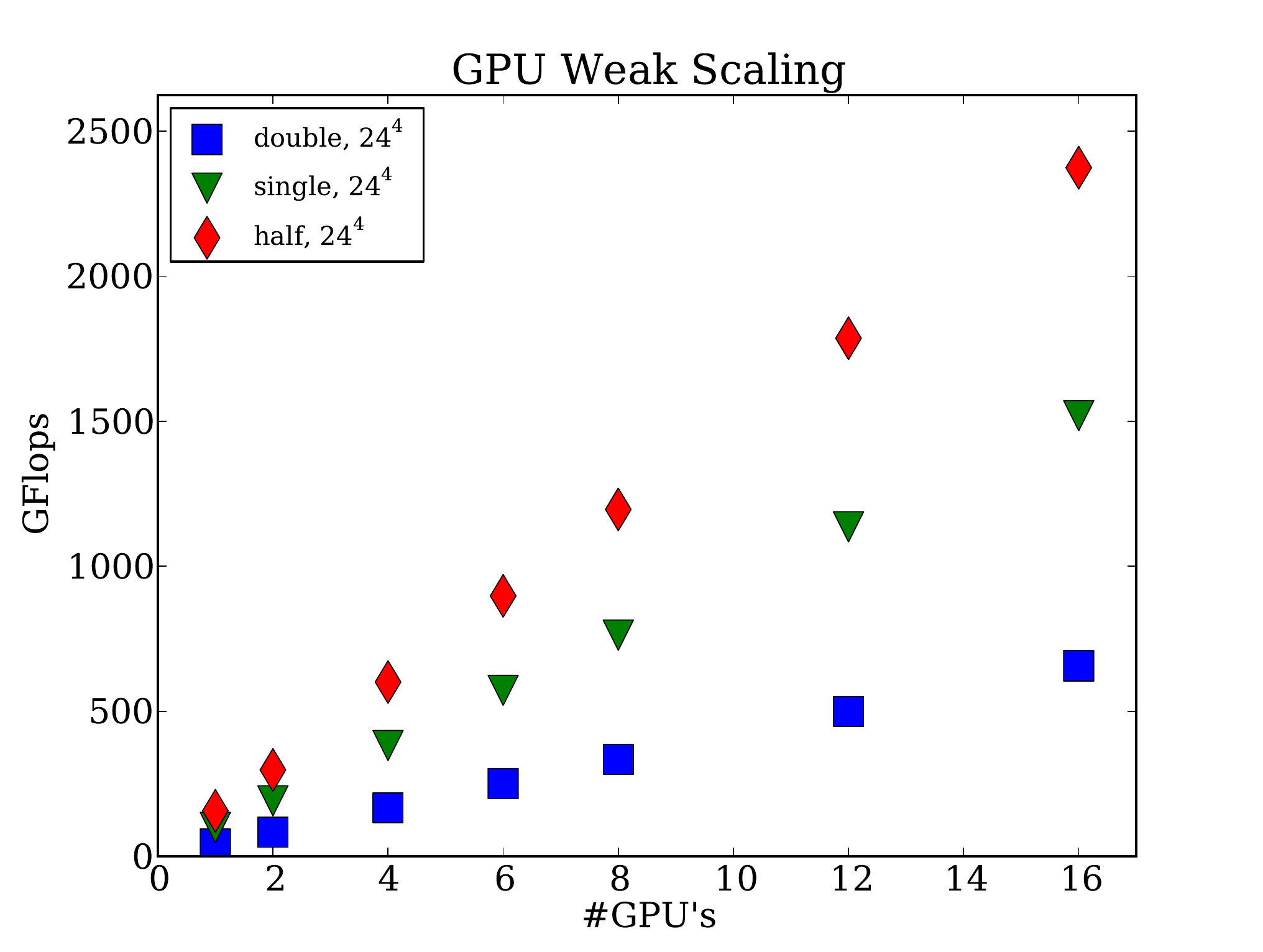} &
\includegraphics[width=0.42\textwidth,angle=0]{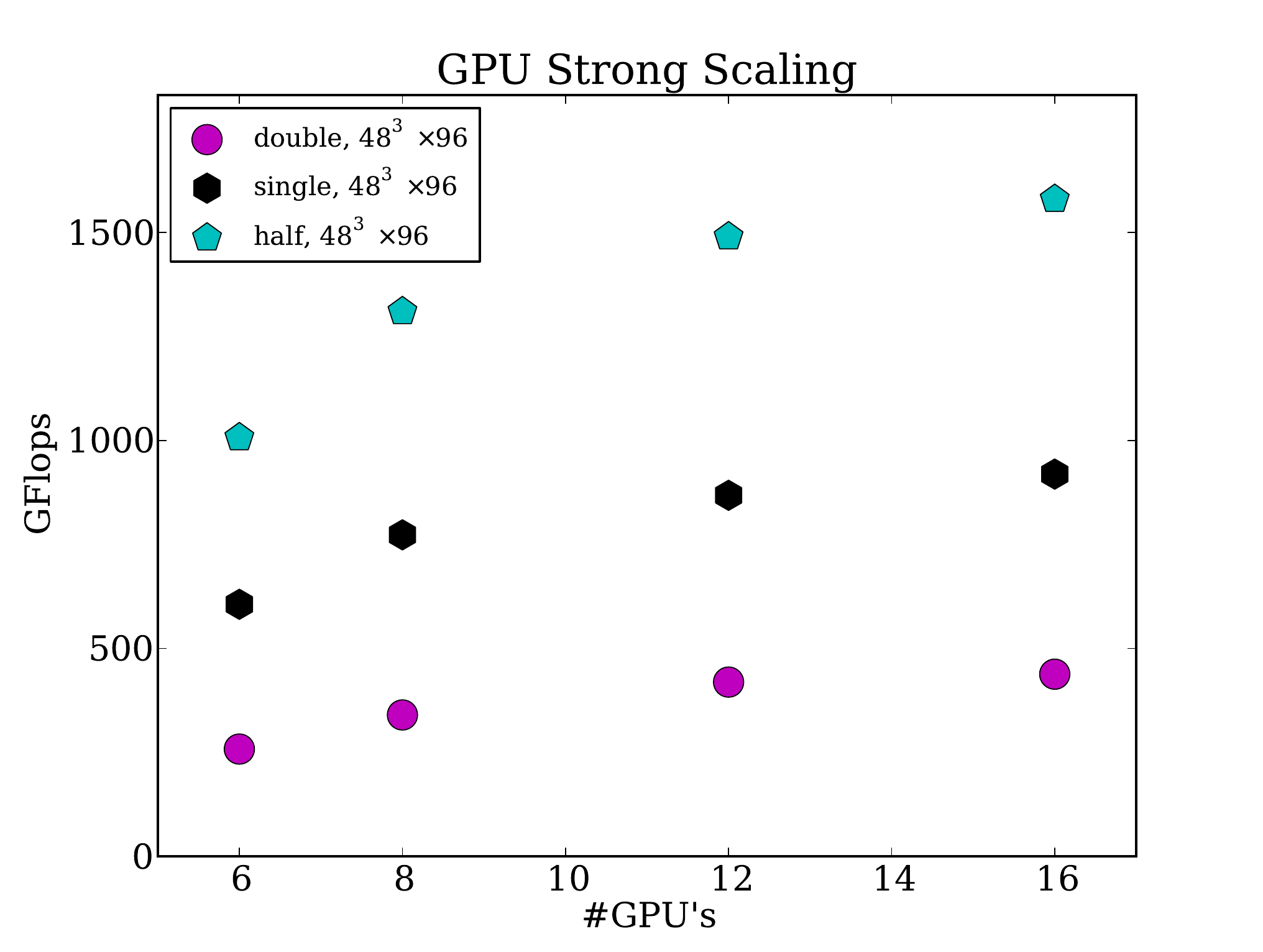}
\end{array}$
\caption{GPU scaling performance of the twisted mass CG solver of the QUDA library in Tesla m2070 cards. Each
node of the cluster featured two GPUs, and the nodes were interconnected with infiniband.\label{Sc}}
\end{center}
\end{figure}

The first variation reduction technique we implemented on GPUs is the Truncated Solver Method (TSM) \cite{TSM}.
In fact, the TSM was our main motivation to move from CPUs to GPUs our calculations of disconnected
diagrams. The reason behind this moving is the high efficiency of the GPUs when dealing with low
precision numbers: in current GPU architectures, computer power is bound to memory bandwidth. Therefore
a reduction in the precision (and consequently in the size of the data read/written to/from memory)
results in an immediate increase of the flop count. In fig.~\ref{Sc} it is shown how the computer power
roughly halves as the size of numerical data is duplicated, from half to single, and from single to
double.

The TSM relies on a low precision estimation of the inverse of the fermionic matrix with high
statistics, corrected by a high precision computation at low statistics. For the low precision
estimation we use around $\sim10^2-10^3$ inverted at low precision ($\rho\sim10^{-2}$), for
which half precision numbers are suitable. The correction, in contrast, requires much higher
precision ($\rho\sim10^{-8}$), and half precision solvers will not converge to desired precision.
We need at least a single precision solver, which, as shown in fig.~\ref{Sc}, is not as efficient
as the half precision solver. Nonetheless, we only need to perform a few ($\sim10-100$) high
precision inversions to get a good correction in most cases, therefore most of the time is spent
in the low precision calculation, which is the one with improved speed.

\subsubsection{The problem of storage}

When the TSM is introduced, one has to face the problem of storage and take decisions regarding which
information can be discarded, due to the large number of stochastic sources generated. In our case
it was decided that only contractions would be stored, but even in this case a storage problem might
appear.

For our calculations, we mainly used the one-end trick for twisted mass fermions (recently included in QUDA
 \cite{tmQ}), which requires volume sources and gives results for all time-slices. We computed all
ultra-local and one-derivative insertions as well, with and without a flavor $\tau_3$ matrix, that is, 160
insertions in total. We also calculated the data for several values of the momenta, up to $p^2 <= 9$. In the
end, each configuration with volume $32^3\times 64$ was taking $\sim50$Gb storage, a monster number taking
into account that GPU application usually are not granted as much disk storage as CPU's. In the GPU clusters
we run our code, we were granted between $5$ and $20$Tb of disk space, which we would fill with 100-400
configurations, a number that might hardly be enough for a single ensemble, let alone when dealing with
several ensembles.

An obvious way to reduce storage needs is to transform the data from text to binary format, which
will grant us a $\sim70\%$ reduction in disk usage, although it will be still a huge amount of data.
A further reduction of storage requirements is only achieved through clever techniques; in our
case we developed a binary-storage technique, inspired in the way the bits make up a byte. This
technique reduced storage requirements up to $\sim97\%$ without losing any relevant information.

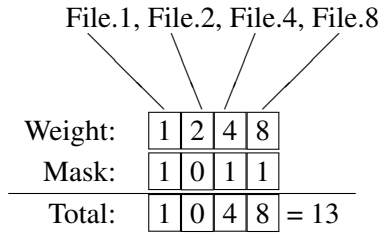
\begin{figure}
\begin{center}
\hspace{3.55cm}File.1, File.2, File.4, File.8
\newline
\setlength{\unitlength}{1cm}
\begin{picture}(8,1)
\put(3.95,1){\line(1,-1){1}}
\put(5.05,1){\line(2,-5){0.4}}
\put(6.15,1){\line(-2,-5){0.4}}
\put(7.25,1){\line(-1,-1){1}}
\end{picture}
\newline
\begin{tabular}{rl}
Weight: & \framebox{1}\framebox{2}\framebox{4}\framebox{8} \\
Mask: & \framebox{1}\framebox{0}\framebox{1}\framebox{1} \\
\hline
Total: & \framebox{1}\framebox{0}\framebox{4}\framebox{8} = 13 \\
\end{tabular}
\caption{Example of construction of the inverse estimator using 13 sources in our storage
method.\label{St}}
\end{center}
\end{figure}

In order to understand the way the technique works, we can have a look at fig.~\ref{St}. The byte
is composed of bits, and each bit has a different wait according to its position (1, 2, 4\ldots).
The idea is to mimic this structure for the stochastic sources. Since in the end we are going
to average the sources, we can add several and store the addition in a single file; so in File.1
file we store the contractions generated with one stochastic source, in File.2 we store the sum
of the contractions coming from the second and the third stochastic source, and so on. Reconstruction
is straightforward, taking into account the base-2 structure.

With this storage method one can recover the data for any number of sources, therefore we are
keeping the same information in much less space. Actually some information is lost, for after
storing the data there is only one way to recover a fixed number of sources, whereas before 
there could be many, but this extra information is not useful for us and can be discarded. In
contrast, we gain a huge reduction in storage requirements, from $O\left(N\right)$ to
$O\left(\log_2 N\right)$.

\subsection{Contraction kernels}

A fundamental step in the calculation of loop amplitudes is the contraction of inverted sources.
To this end we developed efficient GPU code, for the contraction step can be parallelized very easily.
As a result, our contractions are done at speeds of $\sim300$ GFlops in a single Tesla m2070 GPU in
double precision, and the contraction code shows almost perfect scaling with increasing nodes.

Traces were taken in color space, leaving the Dirac and volume indices open. The volume indices are
used later for the FFT, so we obtain solutions for different momenta, whereas the open Dirac indices
are there in order to deal with the different insertions. We calculated the outer product in Dirac
space of both sources to be contracted, and consequently a $4\times4$ matrix was obtained, with enough
information to reconstruct any arbitrary $\gamma$ insertion just by transposition and multiplication.
Therefore, our contraction code automatically outputs all the possible insertions for ultra-local
operators. A covariant-derivative kernel is included, which gives correct results, but in our tests it
was revealed that the contraction performance was not so good when including one-derivative insertions.

In our implementation, we mainly worked with twisted mass fermions and the one-end trick \cite{vvT}, but we
also worked with time-dilution \cite{tDil}. Therefore we developed GPU kernels to contract a single time slice or
a whole vector, returning an array with a time-slice index.

\subsection{Interfaces}

Since QUDA already implements most of the code we need for computing disconnected diagrams, the largest
contribution to the library on this package is the writting of interfaces. Those were designed to calculate
any ultra-local and one-derivative insertion with several variance reduction methods, namely the TSM, the
one-end trick (only for twisted mass fermions), time-dilution and the Hopping Parameter Expansion (HPE) \cite{HPE},
and all the possible combinations of these.

The interface generates random stochastic sources using RANLUX from the GSL library on the CPUs; then the
source is sent to the GPUs for inversion and contraction, and contractions are stored back in the CPUs. This
process is repeated several times for the binary storage system. After we accumulated enough sources, the
data is sent back from CPUs to GPUs for FFT using the nVidia library cuFFT; our output from the last section fits
exactly in the input required in the functions of the cuFFT library, so no further transformations are required.
At this point, all possible momenta are generated, but we usually insert here a cut-off in $p^2$ to reduce storage.
Finally, the results are written to disk in a parallel fashion to reduce I/O time.

\subsection{Limitations and bottlenecks}

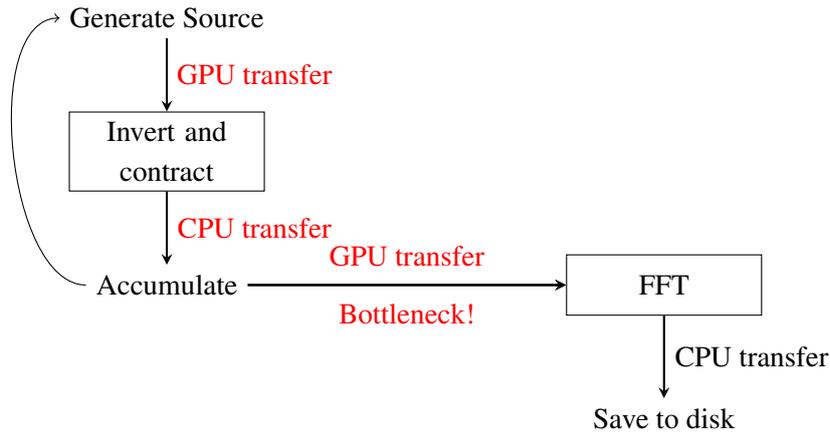
\begin{figure}
  \tikzstyle{deci} = [draw=none, fill=none]
  \tikzstyle{line} = [draw, -stealth, thick]
  \tikzstyle{elli}=[draw, ellipse,minimum height=6mm, text width=12em, text centered]
  \tikzstyle{block} = [draw, rectangle, text width=6em, text centered, minimum height=8mm, node distance=10em]
\begin{center}
  \begin{tikzpicture}
    \node [block] (gpu1) {Invert and contract};
    \node [deci,  above of=gpu1, yshift=2em] (cpu1) {Generate Source};
    \node [deci,  below of=gpu1, yshift=-2em] (cpu2) {Accumulate};
    \node [block, right of=cpu2, xshift=7em] (gpu2) {FFT};
    \node [deci,  below of=gpu2, yshift=-2em](cpu3){Save to disk};

    \path [line] (cpu1) -- node[xshift=3em, color=red] {GPU transfer} (gpu1);
    \path [line] (gpu1) -- node[xshift=3em, color=red] {CPU transfer} (cpu2);
    \path [line] (cpu2) -- node[yshift=1em, color=red] {GPU transfer} (gpu2);
    \path [line] (cpu2) -- node[yshift=-1em, color=red] {Bottleneck!}(gpu2);
    \path [line] (gpu2) -- node[xshift=3em] {CPU transfer} (cpu3);
    \draw [->,out=180,in=180,looseness=0.75] (cpu2.west) to (cpu1.west);

  \end{tikzpicture}
\end{center}
\caption{Implemented scheme in the current version of the discLoop package. The job runs in CPUs, unless
the computation step is inside a box, in that case it runs in GPUs. In red, the detected bottlenecks.\label{bt}}
\end{figure}

Although in our implementation the tough numerical work is done by the GPUs, many operations are still performed
in CPUs, so there is a lot of room for improvement. Memory transfers between CPU and GPU are frequent, and although
these are never done in critical spots of the code, we can yet see performance gains if we move more tasks to the
GPUs. Fig.~\ref{bt} gathers all these problems in a simpel diagram.

\begin{figure}
  \tikzstyle{deci} = [draw=none, fill=none]
  \tikzstyle{line} = [draw, -stealth, thick]
  \tikzstyle{elli}=[draw, ellipse,minimum height=6mm, text width=12em, text centered]
  \tikzstyle{block} = [draw, rectangle, text width=6em, text centered, minimum height=8mm, node distance=4em]
\begin{center}
  \begin{tikzpicture}
    \node [block] (gpu1) {Invert and contract};
    \node [block, above of=gpu1, yshift=2em] (cpu1) {Generate Source};
    \node [block, below of=gpu1, yshift=-2em] (cpu2) {Accumulate};
    \node [block, right of=cpu2, xshift=7em] (gpu2) {FFT};
    \node [deci,  below of=gpu2, yshift=-2em](cpu3){Save to disk};

    \path [line] (cpu1) -- (gpu1);
    \path [line] (gpu1) -- (cpu2);
    \path [line] (cpu2) -- (gpu2);
    \path [line] (gpu2) -- node[xshift=3em] {CPU transfer} (cpu3);
    \draw [->,out=180,in=180,looseness=0.75] (cpu2.west) to (cpu1.west);

  \end{tikzpicture}
\end{center}
\caption{Projected scheme for the package, with most bottlenecks removed and all work done in GPUs.\label{bt2}}
\end{figure}
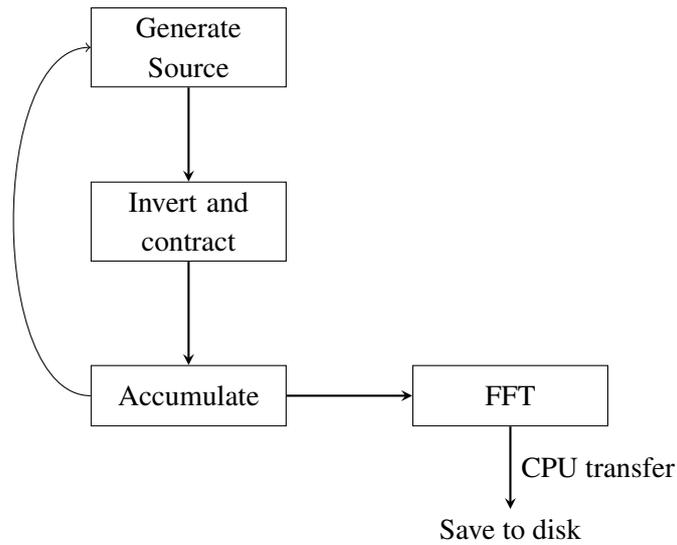

The first bottleneck appears in the generation of the source, we could simply use cuRAND to get random sources
directly on the GPUs. Further on, and after inversion and contraction, the step of accumulation of the sources
introduces a serious overhead coming from memory transfers, mainly when dealing with low precision sources in the TSM,
for the inversion of these sources can take in some scenarios a fraction of a second. Unfortunately, memory constraints
might now allow us to accumulate in GPUs, this mainly depends on the number of operators being computed and the variance
reduction technique used: the one-end trick for instance requires volume sources, and when it is combined with one-
derivative insertions, the amount of data is simply to large to deal with. But other combinations can be easily solved,
like the one-end trick with ultra-local operators or time-dilution with any kind of operator. Our ideal diagram would look
like the one shown in fig.~\ref{bt2}.

Apart from the bottlenecks, the current code suffers from several limitations:
\begin{itemize}
\item Splitting is only allowed in the time-direction, and although splitting in other directions can be implemented,
it is not trivial and will require some time.
\item MultiGPU is currently only supported through MPI, although an update in the code should fix this limitation. From
the QUDA library a huge effort has been done in order to make the communications layer more and more abstract, so these
kind of limitations are removed.
\item Only twisted-mass fermions are supported at this moment. Most likely the package will work with Wilson (and clover) fermions,
but this has not been extensively tested.
\end{itemize}
In spite of these limitations, we are currently using the code in production phase very successfully.

\subsection{Future plans and conclusions}

The successful application of GPUs to the problem of disconnected diagrams allows for an unprecedented precision in
this kind of calculations, due to the huge numerical power available. In our tests, the stochastic error coming from
the fact that the loop amplitudes were being estimated was greatly reduced, or even dissappeared, rendering the
disconnected diagrams accessible. The obstacle that prevented the application of GPUs to this particularly tough
problem was the implementation of efficient code in GPUs, which always is a very difficult task. For that reason,
it is very important to invest in the development of libraries (like QUDA) that can simplify tasks for scientist.
Our package aims to become an easy-to-apply but efficient and powerful solution for computing disconnected diagrams.

The next plans for the package consist of a clean-up and removal of critical limitations, so it can be merged with
master branch of QUDA as soon as possible. Once this feat is accomplished, we can focus in expanding the current
implementation to include other variance reduction techniques, as color- and/or spin-dilution, or in optimizing the
code to increase performance and reduce the bottlenecks. Nonetheless the code as-is can be safely used for production
purposes. It is available at \href{https://github.com/lattice/quda/tree/discLoop}{GitHub} and it is maintained
periodically.

\section*{Acknowledgments}  
Alejandro Vaquero is supported by funding received from the Cyprus RPF under contract
EPYAN/0506/08. This research was in part supported by the Research Executive Agency of the
EU under Grant Agreement number PITN-GA-2009-238353 (ITN STRONGnet) and the infrastructure
project INFRA-2011-1.1.20 number 283286, and the Cyprus RPF under contracts
KY-$\Gamma$A/0310/02 and NEA Y$\Pi$O$\Delta$OMH/$\Sigma$TPATH/0308/31. This work has been
partly supported by the PRACE-1IP and PRACE-2IP (Community Codes Development - Work Package
8) projects funded by the EUs 7th Framework Programme (FP7/2007-2013) under grant agreement
no. RI-211528 and no. RI-283493 respectively, and by SciDAC 2 project. Computer resources
were provided by Cy-Tera at CaSToRC, Forge at NCSA Illinois (USA) and Minotauro at BSC (Spain).

\end{document}